\documentclass[twocolumn,secnumarabic,amssymb,nobibnotes,prb,aps]{revtex4}

\usepackage{graphicx}
\usepackage{dcolumn}
\usepackage{amsmath,amssymb,amsfonts}

\begin{document}
\title{Kinetic Arrest in  Polyion-Induced Inhomogeneously-Charged
 Colloidal Particle Aggregation}

\author{D. Truzzolillo, F. Bordi, F. Sciortino, C. Cametti
 \\
\small\it{Dipartimento di Fisica, Universita' di Roma "La Sapienza"}\\
\small\it{Piazzale A. Moro 5, I-00185 - Rome (Italy) and INFM CRS-SOFT, Unita' di Roma 1}\\
 }

\date{\today}

\begin{abstract}
Polymer chains adsorbed onto oppositely charged spherical
colloidal particles can significantly modify the particle-particle interactions.
For sufficient amounts of added polymers, the original electrostatic repulsion
can even turn into an effective attraction and  relatively large
kinetically stable aggregates  can form.
The attractive interaction contribution between two oppositely
particles arises from the \textit{correlated} adsorption of polyions
at the oppositely charged particle surfaces, resulting in a
non-homogeneous surface charge distribution.  Here, we  investigate
the aggregation kinetics of polyion-induced colloidal complexes
through Monte Carlo simulation, in which the effect of charge
anisotropy is taken into account by a DLVO-like intra-particle
potential, as recentely proposed by Velegol and Thwar [\emph{D.
Velegol and P.K. Thwar, Langmuir, 17, 2001}]. The results reveal
that in the presence of a charge heterogeneity the aggregation
process slows down due to the progressive increase of the potential
barrier height upon clustering. Within this framework, the
experimentally observed cluster phases in polyelectrolyte-liposomes
solutions should be considered as a kinetic arrested state.

\end{abstract}
\maketitle

\section{Introduction}
The addition of oppositely charged polyions to a suspension of
charged colloidal particles gives rise to an intriguing and
partially unexpected phenomenology \cite{bordi1,bordi2}, resulting in the
formation of aggregates which play an important role in a wide
range of implications \cite{napper,felgner1,felgner2}, from membrane biophysics and soft
matter physics \cite{nguyen1,nguyen2,grosberg} to biotechnological processes, such as
therapeutic delivery systems \cite{pedroso}.

These aggregates are governed by a delicate
balance between attractive and repulsive interactions resulting
in the appearance of an equilibrium (or a kinetically arrested)
cluster phase, where single
particles, stuck together by an electrostatic ''glue'' acting between oppositely
charge domains, form reversible, relatively large, complexes.
In this system, the short-range attraction
contribution is promoted by the addition of ''adsorbing''
polyions which form a two-dimensional strongly correlated,
short-range ordered, structure on the surface of the oppositely
charged particle,  contrarily to what happens for
''non-adsorbing'' polymers, where the attraction contribution is
produced by the unbalanced osmotic pressure in the depletion
regime.

In fact, correlated adsorption of polyion chains on the surface
of oppositely charged particles induces two different phenomena:
''charge inversion''\cite{nguyen1,nguyen2,bordi1,bordi2} and
''reentrant condensation''\cite{bordi1,bordi2,bordi3}. The first
one occurs when a colloidal particle binds several oppositely
charged multi-valent ions (polyions), so  that its effective
charge inverts its sign. The second effect, concomitant to the
charge inversion, consists in the formation of particle
aggregates whose average size increases on increasing the
polyion concentration, until it reaches a maximum (at the point
of charge inversion), decreasing afterwards
 to the initial value.

This phenomenology has been investigated in a variety of different colloidal
 systems both for particles bearing a negative charge (such as DOTAP liposomes)
 in the presence of anionic polyions \cite{bordi1} and for positively charged particles
(such as hybrid niosomes) in the presence of cationic polyions\cite{dati}.
\begin{figure}[htbp]
\begin{center}
  \includegraphics[width=9cm]{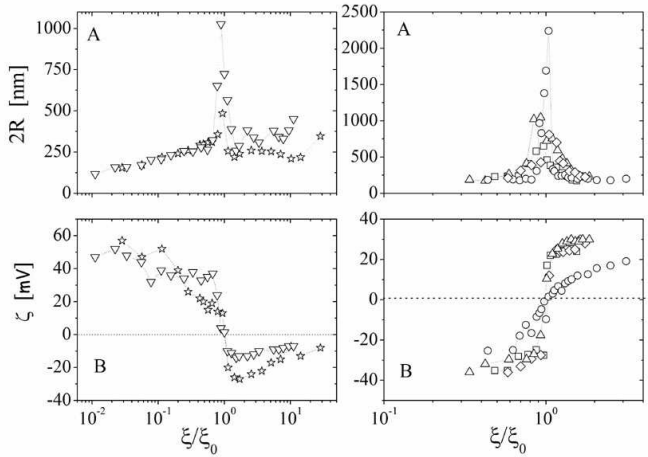}
  \caption{Fig. 1 - A brief  review of the experimental evidence showing a
non-monotonic increase of the size of the polyion-induced
charged particle aggregates, in the presence of oppositely
charged polyions. Both positively and negatively charged
particles are shown. Left panels: reentrant condensation
(aggregate radius $R$) (A) and charge inversion
($\zeta$-potential) (B) of positively charged liposomes, in the
presence of negatively charged polyions. Right panels: reentrant
condensation (aggregate size $R$) (A) and charge inversion
($\zeta$-potential) (B) of negatively charged hybrid niosomes,
in the presence of positively charged polyions. The data are
shown as a function of the charge molar ratio $\xi$ normalized
to the value $\xi_0$ at which the measured $\zeta$-potential
goes to zero. $\xi$ is defined as the polyion to the lipid molar
ratio. The experimental values (hydrodynamic radius measured by
means of dynamic light scattering and $\zeta$-potential measured
by means of Doppler laser electrophoretic techniques) refer to
different colloidal systems. Positive charged particles:
($\bigtriangledown$): DOTAP liposomes (0.8 mg/ml) and
polyacrylate sodium salt; ($\star$): DOTAP liposomes (1.7 mg/ml)
and polyacrylate sodium salt\cite{bordi3}. Negative charge
particles: ($\circ$): hybrid niosomes and $\alpha$-polylysine;
($\bigtriangleup$): hybrid niosomes and $\epsilon$-polylysine;
($\square$): hybrid niosomes and PEVP (ionization degree 65\%);
($\diamond$): hybrid niosomes and PEVP (ionization degree
95\%)\cite{dati}. Hybrid niosomes are built up by Tween20,
Cholesterol and dicethylphosphate and the cationic polyion PEVP
is Poly[N-ethyl-4-vinyl pyridinium]
bromide.}\label{Rz}
\end{center}
\end{figure}
The occurrence of a sharp reentrant condensation on increasing
the adsorption of polyion on charged liposomes has been
documented in a series of articles\cite{bordi1,bordi2,bordi3}.
It has been shown that the size of the aggregates goes through a
maximum and their overall charge inverts its sign in
concomitance to the point of charge inversion, as briefly
reviewed in Fig. \ref{Rz} for some typical cationic or anionic
colloidal particle systems.
The formation of aggregates and, consequently, the stability of
colloidal charged particles (solid particles or soft vesicles)
is usually satisfactorily explained in the framework of
Derjaguin-Landau-Verwey-Overbeek (DLVO) theory \cite{dlvo},
which accounts for electrostatic and van der Waals interactions
between two approaching particles. In the traditional DLVO
theory, the particle surface is assumed to be uniformly charged.
In the case of polyion-correlated adsorption onto the surface of
the particle, this assumption is no longer valid and the usual
DLVO theory might not account for the effects arising from the
surface charge heterogeneity. Indeed, the adsorption of the
polymer generates an ordered charged structure, strongly
dependent on the polyion size and on the valence and linear
charge density \cite{nguyen1}. Some experimental evidences of
such a structure can be found in Refs.[13] and [14].
Transmission electron microscopy [TEM] measurements suggest the
formation of a local ''patchy charge distribution''. In this
respect, it is reasonable to expect that attractive interactions
can arise when "counterion domains" on one particle align  to a
"counterion-free domains" on another approaching particle.

Recently, Velegol and Thwar \cite{velegol} have developed an
analytical model for the effective particle-particle potential
$\langle\Phi\rangle$ in the case of  non-uniform charge
distributions, based on the Derjaguin approximation and on an
extention of  the Hogg-Healy-Fuerstenau [HHF]\cite{hhf} model
for randomly charged surfaces. The resulting spherically
symmetric potential depends on the values of the  average
surface potential (here assumed equal to the $\zeta$-potential)
and on the standard deviation $\sigma$ of the surface potential
among different regions on the particle surface. In particular,
Velegol and Thwar \cite{velegol} showed that the value of the
second moment of the charge distribution lowers the
electrostatic repulsion.

In this work, by means of MC simulation, we analyze the aggregation
process of spherical charged particles interacting via the Velegol
and Thwar \cite{velegol}  DLVO-like potential. Starting from
isolated monomers, we follow the growth of the aggregates until a
kinetically arrested state is reached.  We study the final size of
the aggregates for several values of the $\zeta$-potential and its
standard deviation $\sigma$. The results of the simulations confirm
that the final size of the aggregates grows on increasing  $\sigma$
and on decreasing $\zeta$, providing a microscopic interpretation of
the re-entrant condensation phenomenon of polyion-induced charged
colloidal particle aggregation (Fig.~\ref{Rz}).

\section{Theoretical background}

\begin{figure}[htbp]
\begin{center}
  \includegraphics[width=6cm]{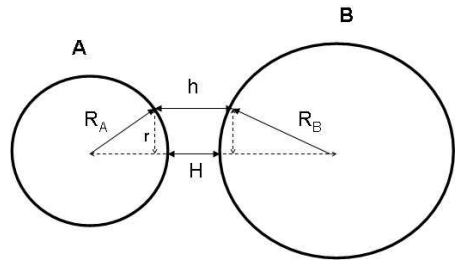}
  \caption{Variables involved in the Derjaguin approximation, for
two approaching particles of radius $R_A$ and $R_B$,
respectively}\label{derj}
\end{center}
\end{figure}

According to the HHF model, the electrostatic potential of mean
force $\Phi$ between two dielectric colloidal spherical
particles of radius $R_A$ and $R_B$ and surface potentials
$\psi_A$ and $\psi_B$, respectively, is given by
$$
\Phi=\frac{\epsilon\pi
R_AR_B}{R_A+R_B}\left[(\psi_A^2+\psi_B^2)\ln(1-e^{-2\kappa H})
\right.$$
\begin{equation}\label{sphere1}
\textstyle +\left.2\psi_A\psi_B\ln\left(\coth\frac{\kappa H}{2}\right)\right]
\end{equation}
Here, $\epsilon$ is the dielectric permittivity of dispersing
medium, $\kappa^{-1}$ is the Debye screening length and $H$ is
the minimal distance between the two particle surfaces (see fig.
\ref{derj}).\\
Eq. (\ref{sphere1}), which is derived under the
conditions $\kappa R_{\alpha}\gg 1$ and $H/R_{\alpha}\ll 1$
($\alpha$=A,B), holds under the following assumption: \emph{i})
$\psi_A$ and $\psi_B$ are relatively small (less than 25 mV at
room temperature); \emph{ii}) the particles share the same
chemical nature and are charged by similar mechanism;
\emph{iii}) the Derjaguin approximation holds\cite{dlvo}.

Velegol and Thwar \cite{velegol} assume that the surface of a
particle $\alpha$ is partitioned in $N$ regions (labelled by
$i$, with $i=1,...,N_{\alpha}$) of area $S$, each of them
characterized by a different value of the  surface potential
$\psi^i_{\alpha}$. These regions must be of sufficient size $L
\sim \sqrt{S}$, so that lateral interactions can be considered
negligible (practically this means that $L\gg \kappa^{-1}$ and
$L\ll H$). The  surface potentials $\psi_A$ and $\psi_B$ of two
approaching particles in the HHF theory are thus replaced by the
random value of the potential $\psi^i_{A}$ and $\psi^j_{B}$,
where $i$ and $j$ label the two domains (on opposite particles)
facing each other. Defining $ \zeta_{\alpha}$ as the average
potential, one can write $\psi^i_{\alpha}  \equiv
\zeta_{\alpha}+\delta^i_{\alpha}$ where $\delta^i_{\alpha}$ a
random contribution that varies among the different regions. The
$\delta^i_{\alpha}$ are independently distributed and not
correlated, i.e.
\begin{equation}\label{ensamble2}
    \langle\psi^i_{\alpha}\psi^j_{\beta}\rangle=\left\{%
\begin{array}{ll}
\zeta_A^2+\sigma_A^2\delta^{ij}, & \hbox{if $\alpha=\beta=A$} \\
\zeta_A\zeta_B, & \hbox{if $A \neq B$} \\
\zeta_B^2+\sigma_B^2\delta^{ij}, & \hbox{if $\alpha=\beta=B$} \\
\end{array}%
\right.
\end{equation}
where $\delta^{ij}$ is the Kr\"{o}enecker delta function and
$\sigma_{\alpha}$ the standard deviation of surface potential on
spheres of type $\alpha$.

The expression for the resulting potential of mean force provided by
Velegol and Thwar \cite{velegol} is
$$
\Phi = \frac{\epsilon\kappa}{2}\sum_{i=1}^M\{(\zeta_A^2+\zeta_B^2+
    2\zeta_A\delta_i^A+2\zeta_B\delta_i^B+\delta_A^2+
    \delta_B^2)(1-\coth \kappa h_i)
$$
\begin{equation}\label{sum2}
+\frac{2(\zeta_A+\delta_i^A)(\zeta_B+\delta_i^B)}{\cosh\kappa
    h_i}\}S
\end{equation}
where the sum runs over the $M$ facing regions and  $h_i$ is the
gap between \emph{i}th regions. The ensemble averaging gives:
\begin{equation}\label{sum3}
 \langle\Phi\rangle=\frac{\epsilon\kappa}{2}\sum_{i=1}^M\{(\zeta_A^2+\zeta_B^2+\sigma_A^2+\sigma_B^2)(1-\coth
\kappa h_i)+ \frac{2\zeta_A\zeta_B}{\cosh\kappa h_i}\}S
\end{equation}
In order to evaluate the sum in Eq. (\ref{sum3}), the Derjaguin
approximation is applied (Fig. \ref{derj}), i.e.
\begin{equation}\label{sum4}
h\approx H+\frac{R_A+R_B}{2R_AR_B}r^2
\end{equation}
and, for any function $F(h)$, the following expression are employed
\begin{equation}\label{sum5}
 \sum_{i=1}^MF(h_i)A_i\approx
 2\pi\frac{R_AR_B}{R_A+R_B}\int_0^{+\infty}F(h)dh
\end{equation}
where $r$ is the flat ring radius (see Fig.~\ref{derj}).\\
The resulting mean force pair interaction potential is given by
$$
\langle\Phi\rangle=\frac{\epsilon\pi
R_AR_B}{R_A+R_B}\left[(\zeta_A^2+\zeta_B^2+\sigma_A^2+\sigma_B^2)\ln(1-e^{-2\kappa
H})+\right.
$$
\begin{equation}\label{finalform}
\left.2\zeta_A\zeta_B\ln\left(\coth\frac{\kappa
H}{2}\right)\right]
\end{equation}

\begin{figure}[htbp]
\begin{center}
  \includegraphics[width=9cm]{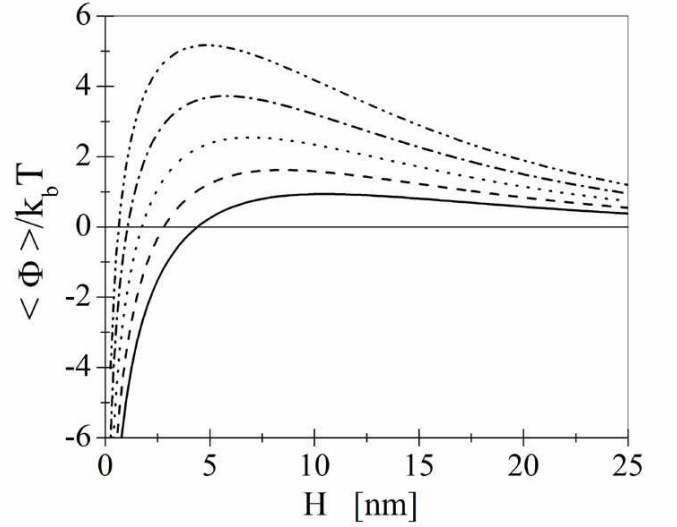}
  \caption{Mean force potential profiles between two identical
spheres calculated from Eq. (\ref{finalform}) for different
values of $\zeta_A=\zeta_B=\zeta$ (from 11 to 19 mV) and for a
constant value of $\sigma_A=\sigma_B=\sigma$=15 mV. Curves are
plotted in units of the thermal energy $k_bT$ at room
temperature for $R_A=R_B=R=40$ nm and $\kappa^{-1}=10$ nm. Solid
line: $\zeta=11$ mV; dashed line: $\zeta=13$ mV; dotted line:
$\zeta=15$ mV; dot-dashed line: $\zeta=17$ mV; dot-dot-dashed
line: $\zeta=19$ mV.}\label{zvar}
\end{center}
\end{figure}

\begin{figure}[htbp]
\begin{center}
  \includegraphics[width=9cm]{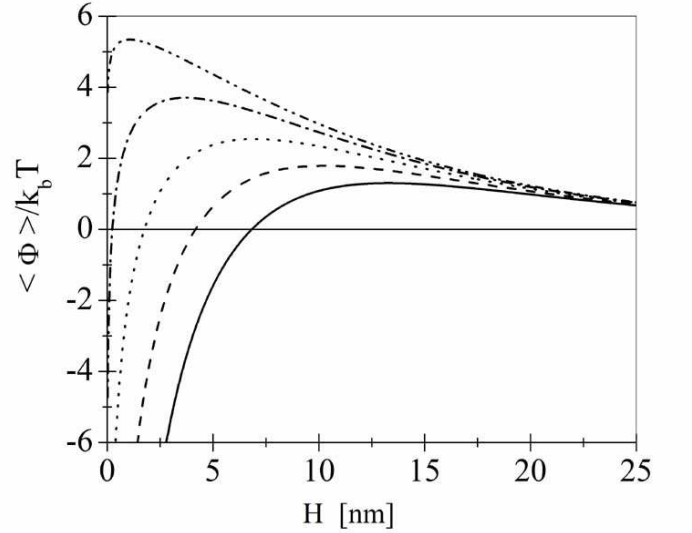}
  \caption{Mean force potential profiles between two identical
spheres calculated from Eq. (\ref{finalform}) for different
values of $\sigma_A=\sigma_B=\sigma$ (from 5 to 25 mV) and for a
constant value of $\zeta_A=\zeta_B=\zeta$=15 mV. Curves are
plotted in units of the thermal energy $k_bT$ at room
temperature for $R_A=R_B=R=40$ nm and $\kappa^{-1}=10$ nm. Solid
line: $\sigma=25$ mV; dashed line: $\sigma=20$ mV; dotted line:
$\sigma=15$ mV; dot-dashed line: $\sigma=10$ mV; dot-dot-dashed
line: $\sigma=5$ mV.}\label{svar}
\end{center}
\end{figure}

The mean force potential for typical values of the
$\zeta$-potential and standard deviation $\sigma$ is shown in
Figs. \ref{zvar} and \ref{svar}. The potential combines a
repulsive net charge-dependent monopole term
($\zeta_{\alpha}\neq 0$) and an attractive multipole term
($\sigma_{\alpha}\neq 0$) arising from the presence of random
charge heterogeneity on the particle surface.\\
\\
These two components lead up to a global maximum and to the presence of a
potential barrier. In the case of particles with the same net
charge, an attractive component is always present since the
first term in Eq. (\ref{finalform}) is negative. If the two
identical particles are uniformly charged ($\sigma_{\alpha}=0$),
both the attractive component and the global maximum vanish,
recovering the HHF expression for two identical spheres
(Eq.~\ref{sphere1}).

The height of the potential barrier that two approaching
particles must overcome in order to aggregate and the distance
between the particle surface at which this maximum occurs can be
evaluated from Eq. (\ref{finalform}). For two identical particles ($R_A=R_B=R$) we obtain:
$$
\langle\Phi\rangle_{max}=\pi\epsilon R \left\{ (\zeta^2+\sigma^2)\ln\left[1-\left(\frac{\zeta^2}{\zeta^2+
    \sigma^2}\right)^2\right]+\right.
    $$
\begin{equation}\label{max}
\left.\zeta^2\ln\left[\frac{2\zeta^2+\sigma^2}{\sigma^2}\right]  \right\}
\end{equation}
and
\begin{equation}\label{Hposmax}
H_{max}=\frac{1}{\kappa}\ln\left(\frac{\zeta^2+\sigma^2}{\zeta^2}\right)
\end{equation}

We point out that: \emph{i}) the height of potential barrier does
not depend on Debye screening length $\kappa^{-1}$. Consequently,
addition of simple salt in bulk phase (varying the ionic strength of
the solvent) does not change the strength of the interaction but
only modifies the distance at which this maximum interaction occurs;
\emph{ii}) using Derjaguin approximation, a linear dependence of the
barrier height on the radius $R$ of the particles arises, underlying
that larger particles interact more intensely than smaller ones and
thus aggregation in the presence of large clusters is relatively
inhibited; \emph{iii}) $\langle\Phi\rangle_{max}$ is zero when
$\zeta=0$ and $\lim_{\sigma\rightarrow 0^+}H_{max}=0$, that is
consistent with the statement that there is no attractive component
for uniformly charged particles and that a lower value of $\sigma$
determines a short-ranged attractive interaction.

In the framework of the above stated model, the aggregation
process can be viewed as a thermally activated process strictly
linked to electrostatic parameters (the $\zeta$-potential and
its standard deviation $\sigma$) and to the clusters dimensions.
Colloidal particles with smaller curvature radius, lower
$\zeta$-potential or higher charge anisotropy will be
characterized by a faster aggregation dynamics.  The aggregation
process will slow down due to the progressive increase of the
repulsive barrier on increasing the size of the aggregates.

In the next section, we introduce a MC simulation method to
analyze the aggregation dynamics of spherical clusters and to
evaluate the evolution of the  mean cluster radius for several
values of $\zeta$-potential and standard deviation $\sigma$,
chosen to model the corresponding experimental range of values
observed for polyion-induced liposome cluster aggregation.

In our calculations, the contribution to the attraction associated
to the van der Waals interaction between two approaching particles
is neglected. The reason why such interaction can be safely
neglected arises from the typical range associated to the van der
Waals interaction, which dyes off before $H_{max}$. Indeed, for
liposome particles (aqueous cores of radii $R_1$ and $R_2$,
respectively, covered by phospholipid shell of thickness $d$), the
van der Waals interaction $V_{vdW}$ can be written as \cite{tadmor}
$$
 V_{vdW}=-\frac{AR_1R_2}{6(R_1+R_2)}\left[\frac{1}{H+2d}-\frac{2}{H+d}+\frac{1}{H}
    \right]-
$$
\begin{equation}\label{Ham}
\frac{A}{6}\ln\left[\frac{H(H+2d)}{2(H+d)}\right]
\end{equation}
(where $A$ is the Hamaker constant). For typical values of the
parameters involved, eq. \ref{Ham} results in a range of
attraction which becomes negligible beyond $1 \div 2$ nm.   As
can be seen from Figs.~\ref{zvar} and ~\ref{svar}, this distance
is always smaller than $H_{max}$. Consequently, we neglect the
van der Waals interaction in the irreversible aggregation
process.

\section{Simulation}
We study  a system composed by $N_p=10000$ colloidal
particles of initial diameter $2R=80$ $nm$ in a cubic box of
volume $V$ with packing fraction $\phi=4\pi\rho R^3/3=0.01$
where $\rho=N_p/V$ is the number density. We carry out MC
simulation using local metropolis algorithm at $T=298$ $K$. Particles interact
via a short range potential defined by Eq. (\ref{finalform}). We study
five different pairs of $\zeta$ and $\sigma$ values,
 comparable to  typical values
measured in liposome solutions\cite{dati,bordi3}. We choose $1/\kappa=10$
$nm$. We recall that  the value of screening length  does not modify the barrier height.

Because of the compactness of the formed clusters, aggregation
is modeled as an "oil drop-like process" in which the shape of
the aggregates retains a spherical form. More precisely, when
two approaching particles ($A$ and $B$) overcome the potential
barrier (surface gap distance $H<H_{max}$), they aggregate
forming an unique particle with radius
$R=\left(R_A^3+R_B^3\right)^{1/3}$ and positioned in the center
of mass of two aggregating spheres. We also assume that $S$ (the
size of the uniform potential regions on the particle surface in
the Velegol-Thwar expression) is independent of the cluster
size.  To incorporate a Brownian dynamics, in the MC algorithm,
the $i$th particle is selected with a probability proportional
to $R_0/R_i$, where $R_0$ is the initial radius and $R_i$ is the
radius of the $i$-th aggregate\cite{babu1, babu2, zimm}.  Each
selected aggregate is moved in each direction by a random
quantity uniformly distributed between $\pm 0.2$ nm. Simulations
have been carried out by varying both the $\zeta$-potential and
the variance $\sigma^2$. The $\zeta$-potential values have been
chosen similar to the typical ones measured in different
colloidal systems (see Fig. 1 and Refs. 10-11). On the contrary,
values for $\sigma^2$, in absence of any experimental
indication, have been chosen arbitrarily, within the validity of
the model and in any case within a range of reasonable values.
The aggregation process progressively slows down and simulations
are interrupted when a plateau in the time dependence of the
aggregate average radius or mass is reached.

\section{Results and Discussion}

\begin{figure}[htbp]
\begin{center}
  \includegraphics[width=9cm]{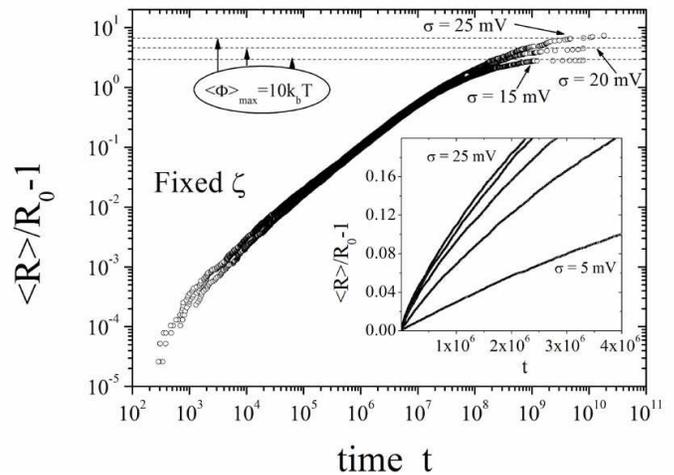}
  \caption{Some typical time evolution of normalized mean cluster
radius $<R>/R_0$-1. Simulations have been carried out for
different values of the standard deviation $\sigma$ (in the
range from $5$ to $25$ mV) with a constant value of the
$\zeta$-potential $\zeta$= $15$ mV. The inset shows, in linear
scale, the evolution at short times of the mean size of the
aggregates, for different values of the standard deviation
$\sigma$ ($\sigma$=25, 20, 15, 10, 5 mV, from top to bottom,
respectively).}\label{logs}
\end{center}
\end{figure}

\begin{figure}[htbp]
\begin{center}
  \includegraphics[width=9cm]{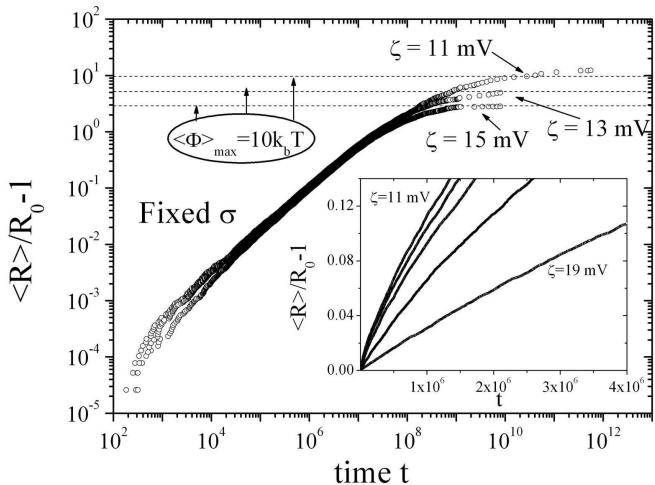}
  \caption{Some typical time evolution of normalized mean cluster
radius $<R>/R_0$-1. Simulations have been carried out for
different values of $\zeta$-potential (in the range from $11$ to
$19$ mV) with a constant value of the standard deviation
$\sigma$= $15$ mV. The inset shows, in a linear scale, the
evolution at short times of the mean size of the aggregates for
different values of the $\zeta$-potential ($\zeta$= 11, 13, 15,
17, 19 mV, from top to bottom, respectively).}\label{logz}
\end{center}
\end{figure}

In Figs. \ref{logz} and \ref{logs}, we present results of the
aggregation process investigated for different values of the
characteristic parameters ($\zeta$-potential and standard deviation
$\sigma$).

First, we will discuss the time evolution of mean cluster radius at
constant $\sigma$ on varying the $\zeta$-potential. Fig.~\ref{logz}
shows that on increasing of the $\zeta$-potential, the aggregation
process slows down and that the aggregates reach at long time a
final limiting size. The slowing down of the dynamics is already
seen at the early stage of the aggregation process, as shown in the
inset of Fig. \ref{logz}, consistent with the effect of
$\zeta$-potential on the height of the potential barrier.

Fig.~\ref{logs} shows the complementary case, i.e, the effect of
the different values of the standard deviation $\sigma$ at a
constant value of the $\zeta$-potential. Again, data confirm
that the increase of the random fluctuations of the charges on
the particle surface speeds up the aggregation process. In all
the cases investigated, the growth process slows down at long
times and the aggregates appear to reach a long-time limit
value, providing a strong evidence for a dynamics slow down and
an arrest on the timescale sampled by our simulation. This
finding suggests that a kinetically arrested state can be
generated in this class of systems by the coupling between the
aggregates size and the electrostatic barrier. An estimate of
the characteristic size reached by the aggregates at long time
can be calculated assuming that the arrest is observed on the
explored time scales when the interaction energy barrier reaches
an \emph{ad-hoc} value. Indeed, inverting Eq.~\ref{max}, the
relation connecting the radius of interacting aggregates and the
potential barrier is easily obtained.
We find that, in all the cases investigated, for all the couples
of $\zeta$-potential and standard deviation $\sigma$ we have
considered, the plateau of the average aggregate size
corresponds to a characteristic barrier $\langle \Phi
\rangle_{max}$ of the order of about $10$ $k_bT$ (see Figs.
\ref{logz} and \ref{logs}, where the normalized size $<R>/R_0$-1
calculated from eq. \ref{max} for different values of the
$\zeta$-potential at fixed $\sigma$ or conversely, for different
values of $\sigma$ at fixed $\zeta$-potential is shown).\\
\\
It is worth nothing that $\zeta$ and $\sigma$ values are
considered to be both constant during the aggregation process
that evolves from the single polyion-decorated liposome to
liposome cluster. This assumption is justified as a consequence
of the local character of the interaction, where only a local
charge (potential) distribution determines the repulsive and
attractive components of the pair potential. The distribution of
the local charge of the elementary unit (single liposome)
surfaces can be considered independent of the size of clusters.

In all the simulations, at short time below the slow down, the
average size of clusters increases according to a power law
$\langle R\rangle/R_0-1\sim at^b $ with a pre-factor that
decreases from $a=7.2 \cdot 10^{-6}$ to $a=1.6 \cdot 10^{-7}$.
At the same time, the exponent $b$ increases from $b=0.70$ to
$0.88$, as the $\zeta$-potential varies from $\zeta=11$ mV to
$\zeta=19$ mV at a constant value of $\sigma=15$ mV. The same
behavior characterizes the $\sigma$ dependence at fixed
$\zeta=15$ mV in the range from $\sigma=25$ mV down to
$\sigma=5$ mV. As expected, the aggregation process does not
follow the diffusion limited cluster aggregation (DLCA model
\cite{witten}) due to the presence of significant potential
barriers controlling the rate of aggregation. Deviation from
DLCA is also evidenced by the not negligible potential barrier
height at which the arrested state occurs (of order of 10
$k_bT$).

\begin{figure}[htbp]
\begin{center}
  \includegraphics[width=9cm]{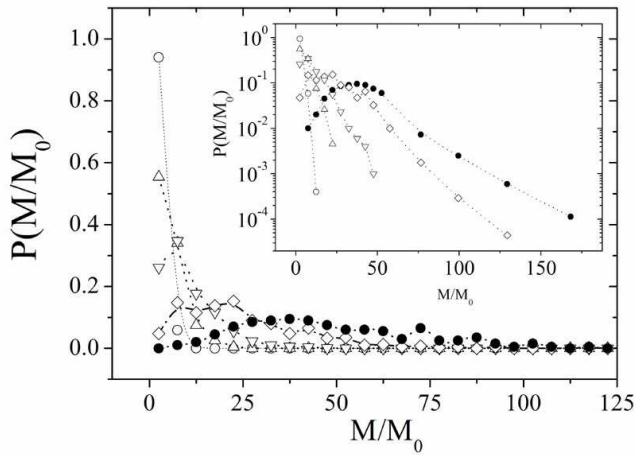}
  \caption{Cluster size distribution of the mass $M$ of the
aggregating particles normalized to the mass $M_0$ of the
initial particles at different aggregation states for $\zeta=$
15 mV and $\sigma=$ 15 mV. ($\circ$): $\langle M\rangle/M_0=2$;
($\bigtriangleup$): $\langle M\rangle/M_0=5$;
($\bigtriangledown$): $\langle M\rangle/M_0=10$;
 ($\diamondsuit$): $\langle M\rangle/M_0=25$; ($\bullet$): $\langle M\rangle/M_0=50$. Inset:
the same distributions in a semi-log scale.}\label{sizez15s15}
\end{center}
\end{figure}

\begin{figure}[htbp]
\begin{center}
  \includegraphics[width=9cm]{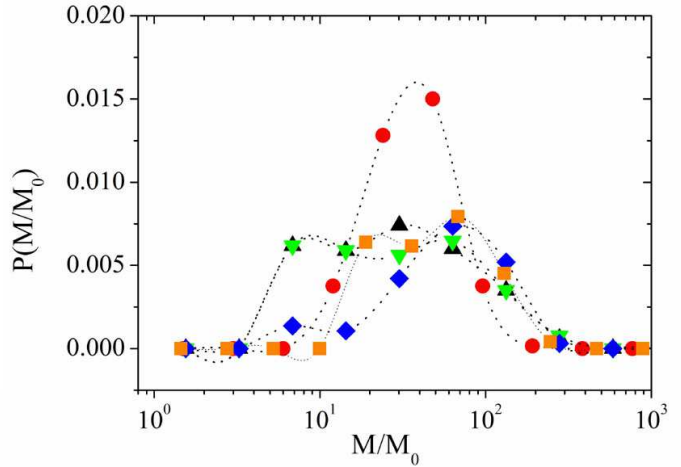}
  \caption{(Color on line) Cluster mass distribution in a
log-scale for the cluster distributions closest to the arrested
state; ($\bullet$): $\zeta$=11 mV, $\sigma$=15 mV;
($\blacktriangle$): $\zeta$=15 mV, $\sigma$ = 15 mV;
($\blacktriangledown$): $\zeta$=15 mV, $\sigma$ = 25 mV;
($\blacklozenge$): $\zeta$=15 mV, $\sigma$ = 20 mV;
($\blacksquare$): $\zeta$=13 mV, $\sigma$ = 15 mV .}\label{fig:final}
\end{center}
\end{figure}

Fig. \ref{sizez15s15} shows the cluster mass distributions at
different stages, during the aggregation process for the case of
the couple of parameters $\zeta=$ $15$ mV and $\sigma=$ $15$ mV.
The system evolves from the initial mono-disperse distribution
toward a broader distribution characterized by a well defined
peak, whose position increases with time. The distribution
freezes when the system kinetically arrests. The inset in Fig.
\ref{sizez15s15} shows the same distribution in a semi-log scale
to highlight the presence of a distribution tail compatible with
an exponential distribution. Similar behavior is observed for
all the cases investigated. The main differences concern the
shape of the distributions in the arrested states, some of which
shown in Fig.~\ref{fig:final} (see also Table~\ref{table:one}).
\begin{table}
  \centering
  \begin{tabular}{c c c}
$\zeta$ [mV]&$\sigma$ [mV]&$\sigma_M$\\
\hline \hline
$11$&$15$&45.3\\
$13$&$15$&38.7\\
$15$&$15$&25.9\\
$15$ &$20$&34.4\\
$15$ &$25$&39.9\\
\hline \hline
\end{tabular}
  \caption{\small{Standard deviation $\sigma_M$ of mass distributions when $\langle M\rangle/M_0=50$
  for the different electrostatic parameters ($\zeta$-potential and its standard
  deviation $\sigma$) used in simulations. }}\label{table:one}
\end{table}
The final size distribution has a smaller standard deviation
$\sigma_M$ in the case of higher values of the $\zeta$-potential
or lower values of $\sigma$. The average mass instead shows an
opposite trend, i.e., it increases for smaller values of the
$\zeta$-potential or higher values of $\sigma$.

\begin{figure}[htbp]
\begin{center}
  \includegraphics[width=9cm]{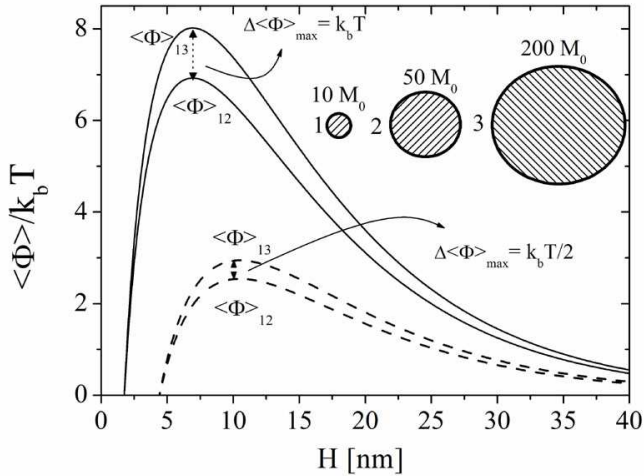}
  \caption{The influence of the particle size on the interaction
potential $\langle \Phi\rangle$. The graph shows the case of a
particle "1" (with mass 10$M_0$) interacting with particle "2"
(with mass 50$M_0$) and particle "3" (with mass 200$M_0$),
respectively, for two different values of the $\zeta$-potential:
$\zeta$=$11$ mV (dashed lines), $\zeta$=$15$ mV (full lines). As
can be seen, particle "1" preferentially aggregates with
particle "2" rather than with particle "3", to which corresponds
a higher potential barrier.}\label{pot-polydisp}
\end{center}
\end{figure}

A possible explanation of the influence of the electrostatic
parameters, $\zeta$-potential and standard deviation $\sigma$,
on the variance $\sigma_M$ of the cluster size distribution is
provided in Fig.~\ref{pot-polydisp}. Here, we compare the
potential profiles (eq. \ref{finalform}) where a particle
interacts with particles of different sizes, for two different
values of the $\zeta$-potential (11 and 15 mV, respectively).
The results show that there is a preferential aggregation which
favors the increase of smaller cluster rather than the cluster
of larger size. Indeed, the inter-particle potential (Eq.
(\ref{finalform})) predicts that the test particle (particle of
mass 10$M_0$ in the example sketched in Fig. \ref{pot-polydisp})
preferentially interact with clusters of smaller size (particle
of mass 50$M_0$) rather than with clusters of larger size
(particle of mass 200$M_0$)
This kind of preferential interaction, which
becomes relatively more and more favorable for larger values of
$\zeta$, favors the aggregation of the smaller clusters in the
system. These privileged interactions result in the formation of
aggregate size distributions whose width decreases with
increasing $\zeta$ or decreasing $\sigma$, as shown in Fig.
\ref{fig:final}.

\section{Comparison with the experimental data}

\begin{figure}[htbp]
\begin{center}
  \includegraphics[width=9cm]{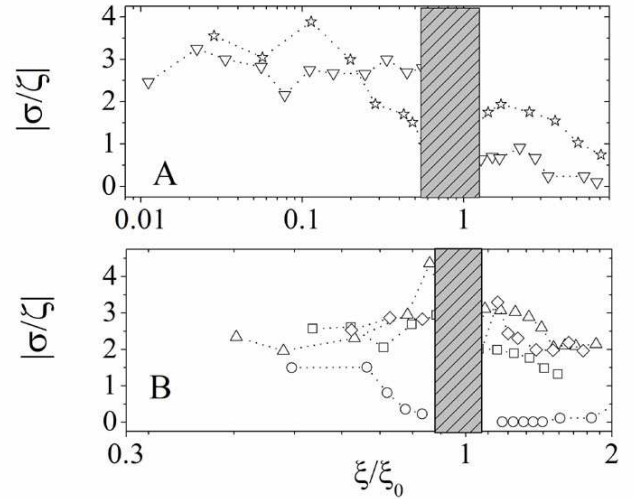}
  \caption{The ratio $|\sigma$/$\zeta|$ calculated from eq. 8 on
the basis of the experimental values of radius $R$ and the
$\zeta$-potential of the aggregates formed in different
polyion-induced particle aggregations. The marked regions
correspond to the experimental instability of the aggregates,
where their average size increases with time, until, in long
time limit, they flocculate. A: Positive charged particles:
($\bigtriangledown$): DOTAP liposomes (0.8 mg/ml) and
polyacrylate sodium salt; ($\star$): DOTAP liposomes (1.7 mg/ml)
and polyacrylate sodium salt\cite{bordi3}. B: Negative charged
particles: ($\circ$): hybrid niosomes and $\alpha$-polylysine;
($\bigtriangleup$): hybrid niosomes and $\epsilon$-polylysine;
($\square$): hybrid niosomes and PEVP (ionization degree 65\%);
($\diamond$): hybrid niosomes and PEVP (ionization degree
95\%)\cite{dati}. Hybrid niosomes are built up by Tween20,
Cholesterol and dicethylphosphate and the cationic polyion PEVP
is Poly[N-ethyl-4-vinyl pyridinium] bromide.}\label{Rzs}
\end{center}
\end{figure}

Eq. 8 provides a connection between the experimentally measured
values of the radius $R$ and the $\zeta$-potential of the
cluster aggregates and the unknown value of the variance
$\sigma^2$ of the surface potential $\psi$. This quantity can be
evaluated for the different systems shown in Fig.~\ref{Rz},
i.e., cationic particles in the presence of anionic polyions
and, conversely, anionic particles in the presence of cationic
polyions. Fig. \ref{Rzs} shows the dependence of the standard
deviation $\sigma$ normalized to the value of the
$\zeta$-potential, $|\sigma/\zeta|$, as a function of the
normalized molar charge ratio $\xi/\xi_0$. Here, $\xi$ is
defined as the ratio between the polyion and lipid molar
concentrations and $\xi_0$ is the value of $xi$ at which the
$\zeta$-potential goes to zero. As can be seen, there is a more
or less pronounced decrease of $|\sigma/\zeta|$ as a function of
charge ratio $\xi/\xi_0$ for all the systems investigated, as
one can expect when more and more polyion adsorption results in
a more uniform charge distribution.

A final comment is in order. Experimentally, in some of the
colloidal systems investigated, it has been observed that, close
to the point of charge inversion, the aggregates do not reach an
equilibrium radius, but their size continues to increase with
time, until, in long time limit, aggregates flocculate. This
window of $\xi/\xi_0$ values is indicated as a dashed area in
Fig.~\ref{Rzs}. Interestingly, if the value of the aggregate
radius  measured just before flocculation is used in Eq.
\ref{max} in conjunction with the corresponding
$\zeta$-potential value, no physical solution for $\sigma$ is
recovered. More precisely, the region of $R$ and $\zeta$ values
for which no physical solutions for $\sigma$ exist is delimited
by the curve $\sigma=0$. Along
 this curve, the equilibrium radius depends on $\zeta$-potential according to
\begin{equation}\label{reqlim}
R=\frac{10k_bT}{\pi\epsilon\zeta^2\ln 4}
\end{equation}.
\begin{figure}[htbp]
\begin{center}
  \includegraphics[width=9cm]{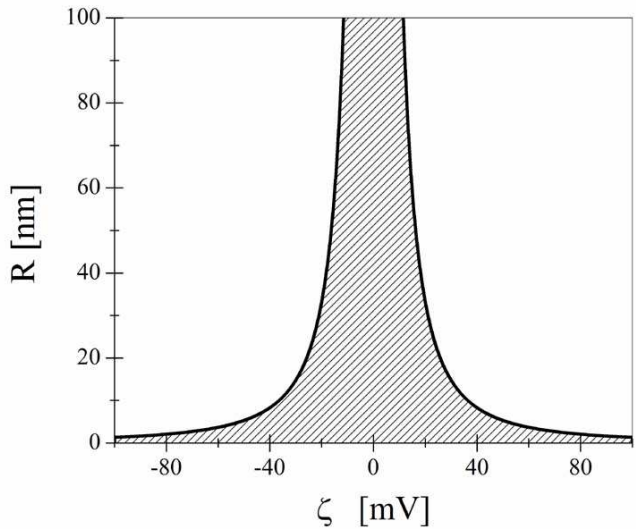}
  \caption{Prohibited region under the curve defined by
expression (\ref{reqlim}) for $T=298$ $K$ and permittivity
$\epsilon$ of water.}\label{instability}
\end{center}
\end{figure}
All the points falling in the dashed windows (instability windows)
of Fig.~\ref{Rzs} are positioned in this area of $R-\zeta$ plane
(dashed area in Fig. \ref{instability}) thus we find particularly
convincing the fact that the Velegol and Thwar potential provides an
estimate of this unstable region because of the close agreement with
the experimental results.

\section{Conclusions}
A large body of experimental \cite{bordi1,bordi2,bordi3},
theoretical \cite{nguyen1,nguyen2,grosberg} and simulation
\cite{messina} investigations have shown that linear flexible or
semiflexible polyions induce the aggregation of oppositely
charged colloidal particles displaying a large variety of
possible structures, thus emphasizing the many facets of charged
macroion complex formation. Among these structures, clusters of
liposomes stuck together by oppositely charged polyions form a
class of model colloids in soft-matter physics which are of the
great importance in many biotechnological implications. In these
cases, the attractive interaction contribution to the
inter-particle potential is originated by a correlated
adsorption at the particle surface, which causes a
non-homogeneous charge distribution at the particle surface.

Since the major driving force of these processes is of
electrostatic origin, more refine interaction potentials should
be used to take into account some peculiar features in the
aggregate cluster formation. We have modeled particle
interactions by means of a mean force inter-particle potential
recently proposed by Velegol and Thwar, who developed a closed
form analytical model to estimate the effect of a non-uniform
charge distribution. Within the scenario of polyion-induced
charged particle aggregation, this potential provides a
justification for why an attractive contribution arises in
like-charged particle interactions in the presence of heterogeneities on the charged surface.

Monte Carlo simulations have been carried out for an ensemble of
particles interacting via a DLVO-like-Velegol and Thwar
potential, for different values of the parameters characterizing
the system. Simulations qualitatively reproduce all
experimentally observable trends in a variety of different
colloidal systems, from positively charged liposomes in the
presence of anionic polyions to negatively charged hybrid
niosomes in the presence of cationic polyions. In particular,
simulations clearly evidence the formation of an arrested
cluster phase, which is the sign of a kinetic arrested state in
low-density colloidal suspensions.

For electrostatically highly coupled systems, the use of the
Velegol and Thwar potential offers an interesting promise for
the understanding of the role of electrostatic attractive and
repulsive interactions between charged colloidal particles and
oppositely charged polyions.


\begin{thebibliography}{22}
\expandafter\ifx\csname
natexlab\endcsname\relax\def\natexlab#1{#1}\fi
\expandafter\ifx\csname bibnamefont\endcsname\relax
  \def\bibnamefont#1{#1}\fi
\expandafter\ifx\csname bibfnamefont\endcsname\relax
  \def\bibfnamefont#1{#1}\fi
\expandafter\ifx\csname citenamefont\endcsname\relax
  \def\citenamefont#1{#1}\fi
\expandafter\ifx\csname url\endcsname\relax
  \def\url#1{\texttt{#1}}\fi
\expandafter\ifx\csname
urlprefix\endcsname\relax\def\urlprefix{URL }\fi
\providecommand{\bibinfo}[2]{#2}
\providecommand{\eprint}[2][]{\url{#2}}

\bibitem[{\citenamefont{Bordi et~al.}(2007)\citenamefont{Bordi, Cametti,
  Sennato, and Viscomi}}]{bordi1}
\bibinfo{author}{\bibfnamefont{F.}~\bibnamefont{Bordi}},
  \bibinfo{author}{\bibfnamefont{C.}~\bibnamefont{Cametti}},
  \bibinfo{author}{\bibfnamefont{S.}~\bibnamefont{Sennato}}, \bibnamefont{and}
  \bibinfo{author}{\bibfnamefont{D.}~\bibnamefont{Viscomi}},
  \bibinfo{journal}{J. Chem. Phys.} \textbf{\bibinfo{volume}{126}},
  \bibinfo{pages}{024902} (\bibinfo{year}{2007}).

\bibitem[{\citenamefont{Bordi et~al.}(2005)\citenamefont{Bordi, Cametti,
  Diociaiuti, and Sennato}}]{bordi2}
\bibinfo{author}{\bibfnamefont{F.}~\bibnamefont{Bordi}},
  \bibinfo{author}{\bibfnamefont{C.}~\bibnamefont{Cametti}},
  \bibinfo{author}{\bibfnamefont{M.}~\bibnamefont{Diociaiuti}},
  \bibnamefont{and} \bibinfo{author}{\bibfnamefont{S.}~\bibnamefont{Sennato}},
  \bibinfo{journal}{Phys. Rev. E R. Comm.} \textbf{\bibinfo{volume}{71}},
  \bibinfo{pages}{050401} (\bibinfo{year}{2005}).

\bibitem[{\citenamefont{Napper}(1983)}]{napper}
\bibinfo{author}{\bibfnamefont{D.}~\bibnamefont{Napper}},
  \emph{\bibinfo{title}{Polymeric Stabilization of Colloidal Dispersion}}
  (\bibinfo{publisher}{Academic}, \bibinfo{address}{London},
  \bibinfo{year}{1983}).

\bibitem[{\citenamefont{Felgner et~al.}(1994)\citenamefont{Felgner, Kumar,
  Sridhar, Wheeler, Tsai, Border, Ramsey, Martin, and Felgner}}]{felgner1}
\bibinfo{author}{\bibfnamefont{J.~H.} \bibnamefont{Felgner}},
  \bibinfo{author}{\bibfnamefont{R.}~\bibnamefont{Kumar}},
  \bibinfo{author}{\bibfnamefont{C.~N.} \bibnamefont{Sridhar}},
  \bibinfo{author}{\bibfnamefont{C.~J.} \bibnamefont{Wheeler}},
  \bibinfo{author}{\bibfnamefont{Y.~J.} \bibnamefont{Tsai}},
  \bibinfo{author}{\bibfnamefont{R.}~\bibnamefont{Border}},
  \bibinfo{author}{\bibfnamefont{P.}~\bibnamefont{Ramsey}},
  \bibinfo{author}{\bibfnamefont{M.}~\bibnamefont{Martin}}, \bibnamefont{and}
  \bibinfo{author}{\bibfnamefont{P.~L.} \bibnamefont{Felgner}},
  \bibinfo{journal}{J. Biol. Chem.} \textbf{\bibinfo{volume}{269}},
  \bibinfo{pages}{2550} (\bibinfo{year}{1994}).

\bibitem[{\citenamefont{Felgner}(1987)}]{felgner2}
\bibinfo{author}{\bibfnamefont{J.}~\bibnamefont{Felgner}},
  \bibinfo{journal}{Proc. Natl. Acad. Sci. USA.} \textbf{\bibinfo{volume}{84}},
  \bibinfo{pages}{7413} (\bibinfo{year}{1987}).

\bibitem[{\citenamefont{Nguyen and Shklovskii}(2001)}]{nguyen1}
\bibinfo{author}{\bibfnamefont{T.~T.} \bibnamefont{Nguyen}} \bibnamefont{and}
  \bibinfo{author}{\bibfnamefont{B.~I.} \bibnamefont{Shklovskii}},
  \bibinfo{journal}{J. Chem. Phys.} \textbf{\bibinfo{volume}{114}},
  \bibinfo{pages}{5905} (\bibinfo{year}{2001}).

\bibitem[{\citenamefont{Nguyen et~al.}(2000)\citenamefont{Nguyen, Grosberg, and
  Shklovskii}}]{nguyen2}
\bibinfo{author}{\bibfnamefont{T.}~\bibnamefont{Nguyen}},
  \bibinfo{author}{\bibfnamefont{A.}~\bibnamefont{Grosberg}}, \bibnamefont{and}
  \bibinfo{author}{\bibfnamefont{B.}~\bibnamefont{Shklovskii}},
  \bibinfo{journal}{Phys Rev. Lett.} \textbf{\bibinfo{volume}{85}},
  \bibinfo{pages}{1568} (\bibinfo{year}{2000}).

\bibitem[{\citenamefont{Grosberg et~al.}(2002)\citenamefont{Grosberg, Nguyen,
  and Shklovskii}}]{grosberg}
\bibinfo{author}{\bibfnamefont{A.~Y.} \bibnamefont{Grosberg}},
  \bibinfo{author}{\bibfnamefont{T.}~\bibnamefont{Nguyen}}, \bibnamefont{and}
  \bibinfo{author}{\bibfnamefont{B.~I.} \bibnamefont{Shklovskii}},
  \bibinfo{journal}{Rev. Mod. Phys.} \textbf{\bibinfo{volume}{74}},
  \bibinfo{pages}{329} (\bibinfo{year}{2002}).

\bibitem[{\citenamefont{Lima et~al.}(2005)\citenamefont{Lima, Simoes, Pires,
  and Faneca}}]{pedroso}
\bibinfo{author}{\bibfnamefont{M.~P.~D.} \bibnamefont{Lima}},
  \bibinfo{author}{\bibfnamefont{S.}~\bibnamefont{Simoes}},
  \bibinfo{author}{\bibfnamefont{P.}~\bibnamefont{Pires}}, \bibnamefont{and}
  \bibinfo{author}{\bibfnamefont{H.}~\bibnamefont{Faneca}},
  \bibinfo{journal}{Phys. Rev. E} \textbf{\bibinfo{volume}{71}},
  \bibinfo{pages}{050401} (\bibinfo{year}{2005}).

\bibitem[{\citenamefont{Bordi et~al.}(2006)\citenamefont{Bordi, Cametti,
  Sennato, and Viscomi}}]{bordi3}
\bibinfo{author}{\bibfnamefont{F.}~\bibnamefont{Bordi}},
  \bibinfo{author}{\bibfnamefont{C.}~\bibnamefont{Cametti}},
  \bibinfo{author}{\bibfnamefont{S.}~\bibnamefont{Sennato}}, \bibnamefont{and}
  \bibinfo{author}{\bibfnamefont{D.}~\bibnamefont{Viscomi}},
  \bibinfo{journal}{Phys. Rev. E R. Comm.} \textbf{\bibinfo{volume}{74}},
  \bibinfo{pages}{030402R} (\bibinfo{year}{2006}).

\bibitem[{\citenamefont{Sennato et~al.}(2007)\citenamefont{Sennato, Bordi,
  Cametti, Marianecci, and Carafa}}]{dati}
\bibinfo{author}{\bibfnamefont{S.}~\bibnamefont{Sennato}},
  \bibinfo{author}{\bibfnamefont{F.}~\bibnamefont{Bordi}},
  \bibinfo{author}{\bibfnamefont{C.}~\bibnamefont{Cametti}},
  \bibinfo{author}{\bibfnamefont{C.}~\bibnamefont{Marianecci}},
  \bibnamefont{and} \bibinfo{author}{\bibfnamefont{M.}~\bibnamefont{Carafa}},
  \bibinfo{journal}{J. Phys. Chem. B,} \textbf{\bibinfo{volume}{}}
  (\bibinfo{year}{2007, 10.1021/jp 0775449}).

\bibitem[{\citenamefont{Verwey and Overbeek}(1948)}]{dlvo}
\bibinfo{author}{\bibfnamefont{E.~J.~W.} \bibnamefont{Verwey}}
  \bibnamefont{and} \bibinfo{author}{\bibfnamefont{J.~T.~G.}
  \bibnamefont{Overbeek}}, \emph{\bibinfo{title}{Theory of the Stability of
  Lyophobic Colloids}} (\bibinfo{publisher}{Elsevier},
  \bibinfo{address}{Amsterdam}, \bibinfo{year}{1948}).

\bibitem[{\citenamefont{Sennato et~al.}(2005)\citenamefont{Sennato, Bordi,
  Cametti, Diociaiuti, and Malaspina}}]{sennatob}
\bibinfo{author}{\bibfnamefont{S.}~\bibnamefont{Sennato}},
  \bibinfo{author}{\bibfnamefont{F.}~\bibnamefont{Bordi}},
  \bibinfo{author}{\bibfnamefont{C.}~\bibnamefont{Cametti}},
  \bibinfo{author}{\bibfnamefont{M.}~\bibnamefont{Diociaiuti}},
  \bibnamefont{and}
  \bibinfo{author}{\bibfnamefont{P.}~\bibnamefont{Malaspina}},
  \bibinfo{journal}{Biochem. Biophys. Acta.} \textbf{\bibinfo{volume}{1714}},
  \bibinfo{pages}{11} (\bibinfo{year}{2005}).

\bibitem[{\citenamefont{Mou et~al.}(1995)\citenamefont{Mou, Czajkowsky, Zhang,
  and Shao}}]{mou}
\bibinfo{author}{\bibfnamefont{J.}~\bibnamefont{Mou}},
  \bibinfo{author}{\bibfnamefont{D.}~\bibnamefont{Czajkowsky}},
  \bibinfo{author}{\bibfnamefont{Y.}~\bibnamefont{Zhang}}, \bibnamefont{and}
  \bibinfo{author}{\bibfnamefont{Z.}~\bibnamefont{Shao}},
  \bibinfo{journal}{FEBS Lett.} \textbf{\bibinfo{volume}{371}},
  \bibinfo{pages}{279} (\bibinfo{year}{1995}).

\bibitem[{\citenamefont{Velegol and K.Thwar}(2001)}]{velegol}
\bibinfo{author}{\bibfnamefont{D.}~\bibnamefont{Velegol}} \bibnamefont{and}
  \bibinfo{author}{\bibfnamefont{P.}~\bibnamefont{K.Thwar}},
  \bibinfo{journal}{Langmuir} \textbf{\bibinfo{volume}{17}},
  \bibinfo{pages}{7687} (\bibinfo{year}{2001}).

\bibitem[{\citenamefont{Hogg et~al.}(1966)\citenamefont{Hogg, Healy., and
  Fuerstenau}}]{hhf}
\bibinfo{author}{\bibfnamefont{R.}~\bibnamefont{Hogg}},
  \bibinfo{author}{\bibfnamefont{T.~W.} \bibnamefont{Healy.}},
  \bibnamefont{and} \bibinfo{author}{\bibfnamefont{D.~W.}
  \bibnamefont{Fuerstenau}}, \bibinfo{journal}{Trans. Faraday Soc.}
  \textbf{\bibinfo{volume}{62}}, \bibinfo{pages}{1638} (\bibinfo{year}{1966}).

\bibitem[{\citenamefont{Tadmor}(2001)}]{tadmor}
\bibinfo{author}{\bibfnamefont{R.}~\bibnamefont{Tadmor}}, \bibinfo{journal}{J.
  Phys: Cond. Mat.} \textbf{\bibinfo{volume}{13}}, \bibinfo{pages}{L195}
  (\bibinfo{year}{2001}).

\bibitem[{\citenamefont{Babu et~al.}(2006{\natexlab{a}})\citenamefont{Babu,
  Rottereau, Nicolai, Gimel, and Durand}}]{babu1}
\bibinfo{author}{\bibfnamefont{S.}~\bibnamefont{Babu}},
  \bibinfo{author}{\bibfnamefont{M.}~\bibnamefont{Rottereau}},
  \bibinfo{author}{\bibfnamefont{T.}~\bibnamefont{Nicolai}},
  \bibinfo{author}{\bibfnamefont{J.~C.} \bibnamefont{Gimel}}, \bibnamefont{and}
  \bibinfo{author}{\bibfnamefont{D.}~\bibnamefont{Durand}},
  \bibinfo{journal}{Eur. Phys. J. E} \textbf{\bibinfo{volume}{19}},
  \bibinfo{pages}{203} (\bibinfo{year}{2006}{\natexlab{a}}).

\bibitem[{\citenamefont{Babu et~al.}(2006{\natexlab{b}})\citenamefont{Babu,
  Rottereau, Gimel, and Nicolai}}]{babu2}
\bibinfo{author}{\bibfnamefont{S.}~\bibnamefont{Babu}},
  \bibinfo{author}{\bibfnamefont{M.}~\bibnamefont{Rottereau}},
  \bibinfo{author}{\bibfnamefont{J.~C.} \bibnamefont{Gimel}}, \bibnamefont{and}
  \bibinfo{author}{\bibfnamefont{T.}~\bibnamefont{Nicolai}},
  \bibinfo{journal}{J. Chem. Phys.} \textbf{\bibinfo{volume}{125}},
  \bibinfo{pages}{184512} (\bibinfo{year}{2006}{\natexlab{b}}).

\bibitem[{\citenamefont{Zimm}(1956)}]{zimm}
\bibinfo{author}{\bibfnamefont{B.~H.} \bibnamefont{Zimm}}, \bibinfo{journal}{J.
  Chem. Phys.} \textbf{\bibinfo{volume}{24}}, \bibinfo{pages}{269}
  (\bibinfo{year}{1956}).

\bibitem[{\citenamefont{Jr and Sander}(1981)}]{witten}
\bibinfo{author}{\bibfnamefont{T.~A.~W.} \bibnamefont{Jr}} \bibnamefont{and}
  \bibinfo{author}{\bibfnamefont{L.~M.} \bibnamefont{Sander}},
  \bibinfo{journal}{Phys. Rev. Lett.} \textbf{\bibinfo{volume}{19}},
  \bibinfo{pages}{1400} (\bibinfo{year}{1981}).

\bibitem[{\citenamefont{Messina et~al.}(2002)\citenamefont{Messina, Holm, and
  Kremer}}]{messina}
\bibinfo{author}{\bibfnamefont{R.}~\bibnamefont{Messina}},
  \bibinfo{author}{\bibfnamefont{C.}~\bibnamefont{Holm}}, \bibnamefont{and}
  \bibinfo{author}{\bibfnamefont{K.}~\bibnamefont{Kremer}},
  \bibinfo{journal}{Phys. Rev. E} \textbf{\bibinfo{volume}{65}},
  \bibinfo{pages}{041805} (\bibinfo{year}{2002}).

\end{thebibliography}
\end{document}